\documentstyle[12pt,epsf]{article}

\newcommand{\be}{\begin{equation}}
\newcommand{\ee}{\end{equation}}
\newcommand{\ba}{\begin{eqnarray}}
\newcommand{\ea}{\end{eqnarray}}
\newcommand{\non}{\nonumber\\}
\newcommand{\nom}{\nonumber}
\newcommand{\del}{\partial}
\newcommand{\dis}{\displaystyle}

\newcommand{\hp}{hep-th/}

\newcommand{\ads}{$AdS_3$}
\newcommand{\ks}{\mathop{k_{\it s\ell}}}

\newcommand{\br}{\mbox{$\bf R$}}

\newcommand{\bz}{\mbox{$\bf Z$}}

\newcommand{\ket}[1]{{|#1 \rangle}}

\newcommand{\cleqn}{\setcounter{equation}{0}}
\newcommand{\en}{\bigcirc}

\newcommand{\bm}[1]{\mbox{\boldmath $#1$}}

\newcommand{\Ncal}{{\cal N}}

\newcommand{\kSL}{k_{s\ell}}
\newcommand{\kSO}{k_{so}}
\newcommand{\ktSO}{\tilde{k}_{so}}

\newcommand{\1}{^{(1)}{}}
\newcommand{\2}{^{(2)}{}}

\newcommand{\II}{I\kern-.11em I}
\newcommand{\psb}{\bm{\psi}}
\newcommand{\GO}{\Gamma_{\Omega}}
\newcommand{\p}{^{\prime}{}}

\newcommand{\Sb}{\bar{S}}
\newcommand{\Lb}{\bar{L}}
\newcommand{\Tb}{\bar{T}}

\newcommand{\Qb}{\bar{Q}}
\newcommand{\Kb}{\bar{K}}
\newcommand{\psib}{\bar{\psi}}
\newcommand{\chib}{\bar{\chi}}
\newcommand{\Psib}{\bar{\Psi}}
\newcommand{\gammab}{\bar{\gamma}}
\newcommand{\sigmao}{\hbox to 1em{\bf\hfil 1\hfil}}
\newcommand{\ct}{\tilde{c}}
\newcommand{\semidirectproduct}%
{{\raise0.037ex\hbox{\rule{0.037em}{1.03ex}}\kern-.17em\times}}
\begin{document}

\begin{titlepage}
\nopagebreak
\begin{flushright}
February 1999\hfill 
KUCP-0129\\
hep-th/9902079
\end{flushright}

\vfill
\begin{center}
{\LARGE $AdS_3/CFT_2$ Correspondence and Space-Time}

~

{\LARGE ${\cal N}=3$ Superconformal Algebra}

\vskip 12mm

{\large S.~Yamaguchi${}^{\dag}$, Y.~Ishimoto${}^{\S}$, 
and K.~Sugiyama${}^{\sharp}$}

\vskip 10mm
${}^{\dag ,\S}$
{\sl Graduate School of Human and Environmental Studies}\\
{\sl Kyoto University, Yoshida-Nihon-Matsu cho}\\
{\sl  Sakyo-ku, Kyoto 606-8501, Japan}\\
${}^{\sharp}$
{\sl Department of Fundamental Sciences}\\
{\sl Faculty of Integrated Human Studies, Kyoto University}\\
{\sl Yoshida-Nihon-Matsu cho, Sakyo-ku, Kyoto 606-8501, Japan}\\
\end{center}
\vfill

\begin{abstract}

We study a Wess-Zumino-Witten model with target space
$AdS_3\times (S^3\times S^3\times S^1)/{\bz}_2$.
This allows us to construct space-time ${\cal N}=3$ superconformal
theories. By combining left-, and right-moving parts 
through a GSO and a ${\bz}_2$ projections,
a new asymmetric $({\cal N},\overline{{\cal N}})=(3,1)$ model
is obtained. It has an extra gauge (affine) $SU(2)$
symmetry in the target space of the type \II A string.
An associated configuration is realized as slantwise
intersecting M5-M2 branes with a ${\bz}_2$-fixed 
plane in the M-theory viewpoint.
\end{abstract}

\vfill 
\noindent
\rule{7cm}{0.5pt}\par
\vskip 1mm
{\small \noindent ${}^{\dag}$  
E-mail :\tt{} yamaguch@phys.h.kyoto-u.ac.jp}\\
{\small \noindent ${}^{\S}$  
E-mail :\tt{} ishimoto@phys.h.kyoto-u.ac.jp}\\
{\small \noindent ${}^{\sharp}$  
E-mail :\tt{} sugiyama@phys.h.kyoto-u.ac.jp}
\end{titlepage}

\section{Introduction}
\cleqn

One of the most interesting dualities is the correspondence between
supergravity theories (SUGRA) on $AdS_{d+1}\times {\cal M}$ and the
$d$-dimensional conformal field theories (CFT). Many intensive
researches are in progress in various dimensional cases.
In an excellent paper by J.~ Maldacena\cite{CFT}, enhancements of the
rigid supersymmetries are proposed in the near-horizon geometries 
on anti-de Sitter (AdS) spaces of the SUGRA.
These theories in the limits 
are conjectured to be identified with boundary conformal
field theories, for an instance, the 4 dimensional
${\cal N}=4$ susy Yang-Mills theory realized on 
a boundary of ${AdS_5}$ in type \II B SUGRA.

Boundary geometries in other dimensions are also believed to have
these enlarged symmetries and a better understanding
of these equivalences is expected to have numerous applications
to strongly coupled gauge theories and could make clear dynamics in
the regions. It might be a realization of the profound problem that
the
dynamics of the supergravity theories are effectively controlled by
large $N$ susy Yang-Mills theory in some limit of large solitonic
charges.
In order to unify various string dualities, it could give us a clue to
formulate some fundamental theory by taking D-branes as fundamental
objects.
For instances, classical solutions in $AdS_4$-,$AdS_7$-backgrounds
are currently investigated in the context of the M-theory. 

But most of the works in this subject are restricted to situations where
the classical low energy approximation of the SUGRA is reliable.
Also turning on Ramond-Ramond (RR) backgrounds makes it difficult
to study these $AdS/CFT$ correspondences.

In contrast, three dimensional case is special in many respects and 
analyses of $AdS_3$ gravity are lifted to those in a stringy level
on this background. It supplies a chance for us to establish
this correspondence in quantum levels.
First, in this case the associated CFT is two dimensional and has
infinite dimensional local symmetries.
That makes it possible to allow us
to  several fruitful results through 
concrete calculations.

Second, string theory on $AdS_3$ can be defined without turning on
RR-fields and should be more amenable to traditional
worldsheet methods. In fact by performing an S-duality on the type \II B
background with RR-fields, we obtain a perturbative string theory with 
only NS-NS fields turned on.
The conformal invariance on the worldsheet of
first quantized superstring leads to a classical solution of SUGRA.
That is to say, the $AdS_3/CFT_2$ correspondence in stringy level
is reduced to an equivalence between the worldsheet CFT and the
space-time (boundary) CFT.
From the point of view of perturbative string theory
on $AdS_3$, Giveon et al.\cite{GKS} have constructed directly superconformal
generators in the space-time CFT in terms of physical vertex operators
of the worldsheet CFT. Recently there are some further developments
along this line\cite{MS,marti,no-go,deBoer,gaida,ito,large,larsen,suga}.

R-symmetries of space-time susy theories are reflected in 
isometries of subspaces of the ${\cal M}$.
Most typical examples are illustrated in 
``$S^3$'', ``$S^3\times S^3$'', ``$S^5$''
cases whose isometry groups are $SU(2)$, $SU(2)\times SU(2)$, 
$SO(6)\cong SU(4)$ respectively.
There are space-time ``small'' ${\cal N}=4$ (2 dim), ``large'' 
${\cal N}=4$ (2 dim), and ${\cal N}=4$ (4dim)
superconformal symmetries associated to them.
When one reduces isometries of these spaces, 
less supersymmetric theories can be obtained in space-time.
In order to carry out this program, a simple useful method is
proposed to divide the subspace in terms of some discrete groups
$\Gamma$\cite{kachru}.
By applying this method to the 5 dimensional case, that is, 
$AdS_5\times S^5/\Gamma$ SUGRA, space-time ${\cal N=}0,1,2,4$ theories
are obtained approximately in the leading order of large N limits.

In general, two dimensional case is better understood than other
higher dimensional analogues. We hope that 
the division method will 
be available in order to construct less (${\cal N}<4$)
superconformal theories for
2 dimensional cases without
approximations, namely, in all orders of the large N expansions.

Motivated with this consideration, we focus on the 
superstring theory on $AdS_3$ background
and 
intend to examine space-time ${\cal N}=3$ superconformal generators
in terms  of a Wess-Zumino-Witten (WZW) model.
Many considerations have been given for (small/large) ${\cal N}=4$
space-time SCFT in $AdS_3$ models\cite{GKS,large,ito,suga}. 
But there still remain 
several uncertain points for space-time less susy models.
Our aim is to develop a concrete applicable method to 
construct less susy models in $AdS_3$ string.
We present ${\cal N}=3$ space-time generators explicitly and
investigate associated brane configurations.

The purpose of this paper is to continue this study 
to establish $AdS_3/CFT_2$ correspondence
in less susy models and to apply it
to some additional examples that are of interest in the different
contexts.

The paper is organized as follows.
In section 2, we shortly review the formulation in GKS's paper and
its possible generalizations for direct product target spaces.
We also explain the Wakimoto's representations\cite{waki} 
of affine Lie algebra currents
and associated WZW models in order to fix our conventions in the
paper. We propose a orbifold-type model in the superstring as a new
class with an $AdS_3$ background. Taking a ${\bz}_2$-division on a
product space ``$S^3\times S^3\times S^1$'', 
we construct space-time ${\cal N}=3$ superconformal theories on its
fixed locus, that is, a diagonal ``$S^3_D$'' space.
It is an extension of the work\cite{GKS} to less space-time susy theories.
By imposing several physical conditions, we can obtain space-time
generators in terms of worldsheet operators in the first quantized
superstring.
The algebra is represented linearly and its global part turns out to
be a super Lie algebra $OSp(3|2)$.
A bosonic part $SL(2;{\br})\times SU_D(2)$ is associated with an
isometry of the sigma model target space $AdS_3\times S^3_D$.
We elaborate some aspects of this model and comment on a proposed 
identification.
We also present modes of space-time ${\cal N}=3$ SCFT generators
towards understandings of the spectra of space-time Fock spaces.
As concrete examples, we analyse chiral primary fields in this algebra
and construct them explicitly. 

Next we combine these left-moving ${\cal N}=3$ SCA and right-moving 
correspondings in the type \II A superstring.
Chiralities of two sides are opposite 
because of a GSO projection 
and the resulting theory turns 
out to have an exotic $({\cal N},\overline{\cal N})=(3,1)$
supersymmetry in space-time. 
In addition to the $\overline{\cal N}=1$ Virasoro symmetry,
we observe that 
the right-side part has an extra $\overline{\cal N}=1$ affine $SU(2)$ 
super Lie algebra associated 
with the isometry of the remaining diagonal $S^3_D$.

In section 3, we introduce a brane configuration whose near horizon
geometry will serve as $AdS_3$ background for string propagation.
The configuration includes $M5$, $M2$ branes and a ${\bz}_2$ fixed plane.
We describe the supergravity solution and its near horizon limit.
In fact by shrinking the radius of a longitudinal circle $S^1$,
the geometry in the M-theory becomes the orbifold-type space with a 
${\bz}_2$-action in the type \II A theory 
explored in section 2. 
The conformal symmetries are realized on a ${\bz}_2$-fixed plane in
this classical solution.
We will also explain relations between radii of the spaces and the levels of 
algebras proposed in \cite{GKS}.

Section 4 is devoted to conclusions and comments. In appendix A, we
collect several conventions for spin operators and cocycle factors.

~


\section{Superstring on 
$AdS_3\times (S^3\times S^3\times S^1)/{\bz}_2$}
\cleqn

The two dimensional CFT has infinite dimensional symmetries and is
investigated in detail. The relations between $AdS_3$ gravity and the
2dim CFT are expected to be understood much deeper than those in other 
dimensions. The isometry group of the $AdS_3$ space is $SO(2,2)$, but
the asymptotic symmetry of them is enhanced to 2dim boundary conformal 
algebra.
This asymptotic symmetry
acts on the background fields evaluated at spatial infinity and leaves 
them invariant. The well-known example is the BTZ black hole\cite{BTZ} in
$AdS_3$
and an associated boundary isometry is generated by 
two commuting Virasoro algebras\cite{BH,BTZ}. 
Its central charge is characterized by 3dim Newton constant $G_3$ and
a radius ${\ell}$ of $AdS_3$ as ${\dis \frac{3\ell}{2G_3}}$\cite{BH,strom}.
Also these relations with CFT are not only established in the 
$AdS_3$ SUGRA, 
but also are lifted to those at the level in string theory propagating 
on \ads . In the context of the latter, asymptotic Virasoro generators
are constructed by worldsheet operators in the RNS formalisms in \cite{GKS}.
Let us review these shortly and consider possible target spaces
allowed by consistency conditions.

\subsection{General case}
First we restrict ourselves to ${\cal N}=1$ WZW models with 
target spaces of the direct product type 
\ba
AdS_3\times \prod_{I=1} G_I\,\,,\,\,\,
AdS_3\cong SL(2;{\br})\,.\nom
\ea
Here $G_I$'s are either compact simple groups or abelian U(1) 
groups. The dimension of each $G_I$ is $d_I$ and its dual Coxeter
number $h^{\vee}_I$.
At the quantum level, ${\cal N}=1$ susy affine Lie algebra is  
realized and associated level is $k_I$ for $G_I$. Then central charge
$c_I$ is calculated as
\ba
c_I =\left(\frac{3}{2}-\frac{h^{\vee}_I}{k_I}\right)d_I\,.\nom
\ea
(We refer $I=0$ to an index associated with $G_0=SL(2;{\br})$ group.)
A constraint is given as a conformal anomaly free condition
that expresses a balance of the central charges amongst ghost parts and
matter parts
\ba
c=\sum_{I=0}\left(\frac{3}{2}-\frac{h^{\vee}_I}{k_I}\right)d_I
=15\,.\label{center}
\ea
There is also a relation about a total dimension
\ba
10 =\sum_{I=0}d_I\,.\label{dim}
\ea
Similarly  
the space-time boundary affine
algebra is expected to have a central charge
$\tilde{c}=6kp$ with $p=\oint \frac{dz}{2\pi i} \gamma^{-1}\del \gamma$.
The $\gamma$ is a scalar appearing in the Wakimoto
representation\cite{waki} 
of the $SL(2;{\br})$ currents.
A level of boundary affine Lie algebra $\tilde{k}_I$ is proportional to 
the level $k_I$ of the associated algebra, $\tilde{k}_I=k_Ip$.
Thus Eqs.(\ref{center})(\ref{dim}) lead us to relations between 
two kinds of central charges
\ba
\frac{6^2}{\tilde{c}}=\sum_{I=1}\frac{h^{\vee}_Id_I}{\tilde{k}_I}
\,\,,\,\,\,\sum_{I=1}d_I =7\,.\nom
\ea
We can write down all possible cases satisfying these conditions:
\ba
\begin{array}{llcl}
1. & SL(2;{\br})\times U(1)^7 & \leftrightarrow & AdS_3\times T^7\\
2. & SL(2;{\br})\times SU(2)\times U(1)^4 
& \leftrightarrow & AdS_3\times S^3\times T^4\\
3. & SL(2;{\br})\times SU(2)\times SU(2)\times U(1)
& \leftrightarrow & AdS_3\times S^3\times S^3\times S^1\,.
\end{array}\nom
\ea
In the case $1.$, the level of $SL(2;{\br})$ algebra is infinite
($k=0$)
and the radius of {\ads} blows up. It is a decompactifying limit of {\ads} 
and corresponds to a flat space-time.

The model in case $2.$ corresponds to an elaborated near horizon limit 
of the D1-D5 system. The ``small'' ${\cal N}=4$ CFT is
realized as space-time asymptotic symmetry.
The $AdS_3/CFT_2$ correspondence is investigated
in the framework of type \II B superstring in \cite{GKS}.
Here (small) ${\cal N}=4$ superconformal generators are constructed
from the worldsheet vertex operators.

Similarly a ``large'' ${\cal N}=4$ SCFT is constructed in the model 
in case $3.$\cite{large}. 
The associated space-time central charge $\tilde{c}$ is
expressed by levels $\tilde{k}_1$, $\tilde{k}_2$ of two $SU(2)$s
\ba
\tilde{c}=\frac{6\tilde{k}_1\tilde{k}_2}{\tilde{k}_1+\tilde{k}_2}\,,\nom
\ea
as required in a unitary ${\cal N}=4$ theory.

In the context of worldsheet theory, we have other types of possible models.
In the following, 
we concentrate on a orbifold-type case and analyse its associated
space-time superconformal theory.

\subsection{$\Ncal=3$ Superconformal algebra}
In this paper we propose a orbifold type model, 
that is, a superstring theory on the
$AdS_3\times (S^3\times S^3\times S^1)/{\bz}_2$ 
as a new class with an {\ads} background.
We explore it and construct a space-time ${\cal N}=3$
superconformal algebra.

The string metric of the {\ads}-part is expressed with polar coordinates
$(\phi ,\gamma ,\bar{\gamma})$
\ba
ds^2=\ell^2(d\phi^2+e^{2\phi}d\gamma d\bar{\gamma})\,,\nom
\ea
where $\ell$ is the radius of the {\ads} space and can be regarded as
NS5 brane charges.
This model also has non-vanishing NS-NS 2-form $B_{NS}$ background
\ba
B_{NS}=\ell^2 e^{2\phi}d\gamma \wedge d\bar{\gamma}\,.\nom
\ea
In the WZW model, we can incorporate this metric and the $B_{NS}$
into a worldsheet Lagrangian in the
quantum corrected form.
The bosonic part of the theory can be written
with scalars $\phi$, $\gamma$ and
auxiliary fields $\beta$,
$\bar{\beta}$ with worldsheet spins 1
\ba
&&{\cal L}=\del\phi\bar{\del}\phi -\frac{2}{\alpha_{+}}R^{(2)}\phi
+\beta\bar{\del}\bar{\gamma}+\bar{\beta}\del\bar{\gamma}
-\beta\bar{\beta}\exp\left(-\frac{2\phi}{\alpha_{+}}\right)\,,\non
&&\alpha_{+}=\sqrt{2k-4}\,,\,\,\,(k=\ks +2)\,.\nom
\ea
This worldsheet theory has an affine $SL(2;{\br})$ symmetry with
level\footnote{The radius $\ell$ of {\ads} is related to
  the level $k$ of the $SL(2;{\br})$ 
affine algebra $\ell^2 =k\ell^2_s$, 
$(\ell_s =\sqrt{\alpha'})$.} $k=k_{\it s\ell}+2$
and its generators $\{J^a_B\}$ are expressed in the
Wakimoto representation\cite{waki}
\ba
&&J^{-}_B=\beta\,,\non
&&J^{3}_B=\beta\gamma+\frac{\alpha_{+}}{2}\del\phi\,,\non
&&J^{+}_B=\beta\gamma^2+\alpha_{+}\gamma\del \phi +k\del \gamma\,.\nom
\ea
Free fields $(\beta\,,\,\gamma)$ and $\phi$ have non-vanishing short
distance behaviors only in the following cases
\ba
&&\beta (z)\gamma (w)\sim \frac{1}{z-w}\,,\qquad
\phi (z)\phi (w)\sim -\log (z-w)\,.\nom
\ea
Next we prepare a set of free fermions $\{\psi^a\}$ $(a=+,-,3)$
with OPEs
\ba
&&\psi^a(z)\psi^b(w)\sim \frac{\eta^{ab}}{z-w}\,,\non
&&\eta^{+-}=\eta^{-+}=2\,,\,\,\,\eta^{33}=-1\,.\nom
\ea
These fields are combined into a set of affine $SL(2;{\br})$
currents $J^a_F$ represented as
\ba
&&J^a_F (z)=-\frac{i}{2}\varepsilon^a{}_{bc}\psi^b\psi^c\,,
\,\,\,(a=+,-,3)\,,\non
&&\varepsilon^{+-3}=-2i\,.\nom
\ea
They constitute fermionic parts of an ${\cal N}=1$ affine $SL(2;{\br})$ 
super algebra with a level ${\ks}$.
It is generated by a set of total currents $J^a=J^a_B+J^a_F$.

We apply the same recipe to describe two $S^3$ parts of the model by
two affine $SU(2)$ currents.
Bosonic parts are generated by two sets of currents $K^{(I)a}$
$(I=1,2\,;\,a=+,-,3)$
\ba
&&K^{(I)-}_B=\beta^{(I)}\,,\non
&&K^{(I)3}_B=\beta^{(I)}\gamma^{(I)}+
i\sqrt{\frac{k'}{2}}\del\phi^{(I)}\,,\non
&&K^{(I)+}_B=\beta^{(I)}\gamma^{(I)2}
-i\sqrt{2k'}\gamma^{(I)}\del\phi^{(I)}+(k'-2)\del\gamma^{(I)}\,,\non
&&\beta^{(I)}(z)\gamma^{(J)}(w)\sim \frac{\delta^{IJ}}{z-w}\,,\qquad
\phi^{(I)}(z)\phi^{(J)}(w)\sim -\delta^{IJ}\log (z-w)\,.\nom
\ea
In order to construct fermionic parts of the currents $K^{(I)a}_F$,
we prepare two sets of fermions $\chi^{(I)a}$ $(I=1,2\,;\,a=+,-,3)$.
One can combine them into ${\cal N}=1$ affine $SU(2)$ 
currents with level $2$
\ba
&&K^{(I)a}_F=-\frac{i}{2}\varepsilon^a{}_{bc}\chi^{(I)b}\chi^{(I)c}\,,\non
&&\chi^{(I)a}(z)\chi^{(J)b}(w)\sim
\frac{\delta^{IJ}\tilde{\eta}^{ab}}{z-w}\,,\non
&&\tilde{\eta}^{+-}=\tilde{\eta}^{-+}=2\,,\,\,\,
\tilde{\eta}^{33}=+1\,,\,\,\,\varepsilon^{+-3}=-2i\,.\nom
\ea
The total currents for each $S^3$ are defined as
$K^{(I)a}=K^{(I)a}_B+K^{(I)a}_F$ $(I=1,2\,;\,a=+,-,3)$
with a level $k'$.
A set of a boson $Y$ and a fermion $\lambda$ describes a string
propagating on $S^1$ and associated currents are given by $\del Y$ and 
$\lambda$. 

Now let us consider an anomaly free condition, that is, a criticality
condition of string theory. 
The matter parts in this system contain
$SL(2;{\br})_{\ks}$, $SU(2)^{(I)}_{k'}$ and $U(1)$ affine Lie
algebras and their centers must be balanced with those of ghost parts
\ba
15=3\left(\frac{3}{2}+\frac{2}{\ks}\right)
+2\times 3\left(\frac{3}{2}-\frac{2}{k'}\right)+\frac{3}{2}\,.\nom
\ea
It leads to a relation among two levels ${\dis \ks =\frac{k'}{2}}$.
Then the worldsheet theory has a set of ${\cal N}=1$ superconformal
currents $T(z)$ (energy momentum tensor) and $G(z)$
(super stress tensor)
\ba
&&T(z)=\frac{1}{\ks}
\left(\eta_{ab}J^a_BJ^b_B+\frac{1}{2}\sum_{I=1,2}
\tilde{\eta}_{ab}K^{(I)a}_BK^{(I)b}_B
\right)\non
&&\qquad\qquad 
-\frac{1}{2}\left(\eta_{ab}\psi^a\del\psi^b+\sum_{I=1,2}\tilde{\eta}_{ab}
\chi^{(I)a}\del\chi^{(I)b}+\del Y\del Y+\lambda\del\lambda\right)\,,\non
&&G(z)=
\sqrt{\frac{2}{\ks}}
\left(\eta_{ab}\psi^aJ^b_B-\frac{i}{6}\varepsilon_{abc}\psi^a\psi^b\psi^c
\right)
\non
&&\qquad\qquad 
+\sqrt{\frac{1}{\ks}}\sum_{I=1,2}
\left(\tilde{\eta}_{ab}\chi^{(I)a}K^{(I)b}_B
-\frac{i}{6}\varepsilon_{abc}\chi^{(I)a}\chi^{(I)b}\chi^{(I)c}\right)
+i\lambda\del Y
\,.\label{n1}
\ea
Now we shall explain the ${\bz}_2$-action of our model.
In order to leave operators invariant under an exchange of two $SU(2)$s,
we set an action of the ${\bz}_2$ as
\ba
(\,K^{(1)}\,,\,K^{(2)}\,,\,Y\,)\rightarrow 
(\,K^{(2)}\,,\,K^{(1)}\,,\,-Y\,)\,,\nom
\ea
where the $(\,K^{(1)}\,,\,K^{(2)}\,,\,Y\,)$ is a set of coordinates of
$S^3\times S^3\times S^1$.
This operation picks up only diagonal part of the
$SU(2)\times SU(2)$ algebra and an isometry ``$SU(2)\times SU(2)$''
of large ${\cal N}=4$ is reduced to a diagonal $SU_D(2)$. This
turns out to generate an ${\cal N}=3$ superconformal algebra.

First the diagonal $SU_D(2)$ is invariant under the ${\bz}_2$ action and 
an associated affine Lie algebra is constructed by summing two $SU(2)$'s
\ba
K^a=K^{(1)a}+K^{(2)a}\,.\nom
\ea
The worldsheet $SU_D(2)$ generates an affine Lie algebra with a level
${k}_{\it so}=2{k}'=4{k}_{s\ell}$. 
That is to say, the diagonal ``$S^3$'' has a
radius $\ell_s\sqrt{k_{\it so}}=2\ell_s \sqrt{\ks}$, which is twice   
as large as the radius ($\ell_s \sqrt{\ks}$) of the {\ads}.

By using these currents, we can 
introduce mode operators of space-time affine $SU(2)$ currents 
$T^a_m$ in the $0$-picture
\ba
&&T^a_m=\oint \frac{dz}{2\pi i}
\left[
\gamma^m K^a +2m \gamma^m \chi^a \left(
\psi^3-\frac{1}{2}\psi^{-}\gamma -\frac{1}{2}\psi^{+}\gamma^{-1} 
\right)\right]\,,\non
&&\psi^{\pm}=\psi^1\pm i\psi^2\,,\qquad
\chi^a =\frac{1}{\sqrt{2}}(\chi^{(1)a}+\chi^{(2)a})\,,\nom
\ea
or in the $(-1)$-picture
\ba
T^a_m=\sqrt{2\ks}\oint \frac{dz}{2\pi i}
e^{-\varphi}\chi^a\gamma^m\,.\nom
\ea
The $\varphi$ is the bosonized super-reparametrization ghost.
This space-time affine algebra has a level
\ba
\ktSO=2k'p\,,\,\,\,p=\oint \frac{dz}{2\pi i}\gamma^{-1}\del\gamma\,.\nom
\ea
Also modes of a Virasoro operator $L_m$ have the following formula in the
$0$-picture
\ba
 L_m=-\oint \frac{dz}{2\pi i}\left[
(1-m^2)J^3\gamma^m+
\frac{m(m-1)}{2}J^-\gamma^{m+1}+
\frac{m(m+1)}{2}J^+\gamma^{m-1}
\right]\,.\nom
\ea
They satisfy space-time commutation relations with a 
central charge
$\tilde{c}=3k' p$
and two central terms are associated as
${\dis
\tilde{c}=\frac 32 \ktSO}$.

Next the space-time fermionic generators are constructed from spin
operators\cite{FMS} in the worldsheet theory.
We have ten free fermions in the associated super WZW model
with the target space $AdS_3\times S^3\times S^3\times S^1$:
\ba
\left\{
\begin{array}{ccc}
AdS_3 & \cdots & \psi^1\,,\,\,\psi^2\,,\,\,\psi^3\,,\,\,\\
S^3 & \cdots & \chi^{(1)1}\,,\,\,\chi^{(1)2}\,,\,\,\chi^{(1)3}\,,\,\,\\
S^3 & \cdots & \chi^{(2)1}\,,\,\,\chi^{(2)2}\,,\,\,\chi^{(2)3}\,,\,\,\\
S^1 & \cdots & \lambda\,.
\end{array}
\right.
\nom
\ea
Supercharges (spin operators) before the ${\bz}_2$-operation 
belong to the $({\bf 2}\,,\,{\bf 2})$
representation of the $SU(2)\times SU(2)$. They are decomposed
into ${\bf 3}+{\bf 1}$ representations under the $SU_D(2)$.
The ${\bz}_2$-action projects out the singlet and survives the triplet 
fermionic generators. 
In order to pick up ${\bz}_2$-invariant parts, 
we introduce next combinations
\ba
\begin{array}{lcl}
\psb^1 =\psi^1\,,& & \psb^2 =\psi^2\,,\\
\psb^3 =\chi^{(1)1}\,,& & \psb^4 =\chi^{(1)2}\,,\\
\psb^5 =\chi^{(2)1}\,,& & \psb^6 =\chi^{(2)2}\,,\\
\psb^7 =\frac{1}{\sqrt{2}}(\chi^{(1)3}+\chi^{(2)3})\,,
& & \psb^8 =i\psi^3\,,\\
\psb^9 =\frac{1}{\sqrt{2}}(\chi^{(1)3}-\chi^{(2)3})\,,
& & \psb^{10} =\lambda \,,
\end{array}\nom
\ea
and bosonize these fermions 
\ba
\begin{array}{lcl}
\psb^1\pm i\psb^2 =\sqrt{2}e^{\pm iH_1}\,,& 
\psb^3\pm i\psb^4 =\sqrt{2}e^{\pm iH_2}\,,\\
\psb^5\pm i\psb^6 =\sqrt{2}e^{\pm iH_3}\,,& 
\psb^7\pm i\psb^8 =\sqrt{2}e^{\pm iH_4}\,,\\
\psb^9\pm i\psb^{10} =\sqrt{2}e^{\pm iH_5}\,.& 
& 
\end{array}\nom
\ea
Then spin fields $S_\alpha$ are defined in the $(-1/2)$-picture as
\ba
&& S_{\alpha}=\exp\left(\frac i2 \sum_{I}\alpha_I H^I\right)
c_\alpha\,,\non
&&c_\alpha\,;\,\mbox{cocycle factor}\,.\nom
\ea
The $\alpha_I$ $(I=1,2,\cdots ,5)$ takes its value $\pm 1$.
These fields have non-trivial OPEs
\ba
&& e^{-\varphi/2}S_{\alpha}(z) e^{-\varphi/2}S_{\beta}(w)\sim 
\frac{1}{z-w}e^{-\varphi}(C\Gamma^M)_{\alpha\beta}\psb^M\,,\non
&& \psb^M(z) S_{\alpha}(w) \sim \frac{1}{\sqrt 2 (z-w)^{1/2}}
S_{\beta}(\Gamma^M){}^{\beta}{}_{\alpha}\,,\non
&&\mbox{with} \qquad\{\Gamma^M,\Gamma^N\}=2\delta^{MN}\,,\non
&&C\Gamma^MC^{-1}=-(\Gamma^M)^T\,,\,\,\, 
C^T=-C,\;C^2=1\,.\nom
\ea
Here $\Gamma^M$s are $SO(10)$ gamma matrices and the
$C$ is a matrix representation of a charge conjugation operator.
Also several concrete expressions are put in order in an appendix A.
One can introduce a notation $u(z)$ to represent general types
of spinors
\ba
u(z)=\sum_\alpha S_\alpha (z)u^\alpha\,.\nom
\ea
We can express operator product expansions of the $u(z)$ and $\psb^M$
as
\[
 \psb^M(z) u(w) \sim \frac{1}{\sqrt 2 (z-w)^{1/2}}
S_{\beta}(\Gamma^M){}^{\beta}{}_{\alpha}u^{\alpha}.
\]
As operations on the $u^{\alpha}$s,
the operators \(\psb^M\) can be identified with 
the \((1/\sqrt2)\Gamma^M\).

Naively there seems to be possible $32$ spin operators, but
we have to impose several physical conditions on them.
The first is the BRS condition and it acts on the
$u(z)$. An associated part of this is described by an operator
$\Gamma_G$
\ba
&&\Gamma_G=i\Gamma^1\Gamma^2\Gamma^8
+\frac{1}{2}\Gamma^3\Gamma^4(\Gamma^7+\Gamma^9)
+\frac{1}{2}\Gamma^5\Gamma^6(\Gamma^7-\Gamma^9)\,,\non
&&\Gamma_Gu =0\,.\label{cond1}
\ea
The second condition comes from a GSO projection operator $\Gamma^{11}$
\ba
&&\Gamma^{11}=-i\Gamma^1\Gamma^2\cdots \Gamma^{10}\,,\non
&&\Gamma^{11}u=+u\,.\label{cond2}
\ea
One can impose a positive chirality condition on the left-part in the
worldsheet $\Gamma^{11}u=+u$ only if the number of minus signs of $\alpha_I$ 
is even. Namely, only spinor representation is survived in this
projection.
Oppositely a negative chirality condition leads us to pick up 
some sets $\{\alpha_I\}$ with odd number of minus signs. 
It corresponds to a co-spinor representation.

The last constraint is reduced to the ${\bz}_2$ projection
$\Gamma_{\Omega}$.
This ${\bz}_2$ operation acts on fermionic fields as
\ba
&&(\Gamma^3\,,\,\Gamma^5)\rightarrow (\Gamma^5\,,\,\Gamma^3)\,,\non
&&(\Gamma^4\,,\,\Gamma^6)\rightarrow (\Gamma^6\,,\,\Gamma^4)\,,\non
&&(\Gamma^9\,,\,\Gamma^{10})\rightarrow (-\Gamma^9\,,\,-\Gamma^{10})\,,\non
&&\Gamma^M\rightarrow \Gamma^M\,,\,\,\, (M=1,2,7,8)\,,\nom
\ea
and is described by an operator $\Gamma_{\Omega}$
\ba
&&\GO=\frac 12 \Gamma^9 \Gamma^{10}
(\Gamma^3+ \Gamma^5)(\Gamma^4+ \Gamma^6)
\Gamma^3 \Gamma^{4}\Gamma^5 \Gamma^{6}\,,\non
&&\mbox{with}\qquad \GO^2=1\,.\nom
\ea
${\bz}_2$-invariant spinors must be selected out by a
condition
\ba
\GO u=+u\,.\label{cond3}
\ea
Now we make a remark here: the operator $\GO$ picks up a triplet in
the $SU_D(2)$. But we can equivalently project out a triplet but leave 
a singlet invariant by using another ${\bz}_2$-projection
\ba
\widetilde{\GO}=-\GO \,.\nom
\ea
That leads us to obtain an ${\cal N}=1$ superconformal algebra as we will show
in the next subsection.

We obtain physical spinor operators $S^a_r$ $(a=1,2,3,\,;\,
r=\pm 1/2)$ satisfying the above three conditions 
(\ref{cond1}),(\ref{cond2}),(\ref{cond3})
\ba
&&S^+_{+1/2}=S_{+++++}\,,\non
&&S^0_{+1/2}=-\frac{1}{2\sqrt2}(S_{+-+-+}+S_{+-++-}+S_{++--+}-S_{++-+-})
\,,\non
&&S^-_{+1/2}=-S_{+----}\,,\non
&&S^+_{-1/2}=-S_{-++-+}\,,\non
&&S^0_{-1/2}=\frac{1}{2\sqrt2}(S_{--+++}-S_{--+--}+S_{-+-++}+S_{-+---})\,,\non
&&S^-_{-1/2}=-S_{---+-}\,.\nom
\ea
The $\{S^a_r\}$s belong to representation $({\bf 2}\,,\,{\bf 3})$ 
of the $SL(2\,;\,{\br})\times
SU_D(2)$ and the index ``$r$'' corresponds to a doublet of
$SL(2\,;\,{\br})$, the ``$a$'' is associated to a triplet in
$SU_D(2)$.
We can write down full modes of space-time super currents
$Q^a_r$ in the $(-1/2)$-picture
\ba
 &&Q^a_r=A\oint \frac{dz}{2\pi i}e^{-\varphi/2}S^a_r\,,\non
 &&S^a_r=\left(r+\frac12\right)\gamma^{r-1/2}S^a_{+1/2}
-\left(r-\frac12\right)\gamma^{r+1/2}S^a_{-1/2}\,,\non
&&A^2=-i (8\ks)^{1/2}\,,\,\,\,\left(a=1,2,3\,;\,r\in \frac{1}{2}+\bz
\right)\,.\nom
\ea
These currents $L_n$, $T^a_m$, $Q^a_r$ close 
commutation relations together with a spin $1/2$ fermionic current
$\Psi_r$
\ba
&&\Psi_r=A\oint \frac{dz}{2\pi i}e^{-\varphi/2}P_r\,,\non
 &&P_r=-\frac{1}{2\sqrt2}\Bigg[
\gamma^{r+1/2}(-S_{--+++}-S_{--+--}-S_{-+-++}+S_{-+---})\non
&&\qquad\qquad+\gamma^{r-1/2}(-S_{+-+-+}+S_{+-++-}-S_{++--+}-S_{++-+-})
\Bigg]\,.\nom
\ea
These constitute a space-time ${\cal N}=3$ SCA \cite{extend, schwi, miki} with
central charge ${\dis \tilde{c}=\frac{3}{2}\ktSO}$,
$(\kSO=4\ks\,,\,\ktSO=\kSO p)$
\begin{eqnarray*}
 &&[L_m,L_n]=(m-n)L_{m+n}+\frac{\tilde{c}}{12}(m^3-m)\delta_{m+n,0}
\label{eq1}\,,\\
 &&[T^a_m,T^b_n]=i\varepsilon^{ab}{}_cT^c_{m+n}+
\frac{\ktSO}{2}m\delta_{m+n,0}\tilde{\eta}^{ab}\label{eq2}\,,\\
&&[L_m,T^a_n]=-nT^a_{m+n}\label{eq3}\,,\\
&& [T^a_m,Q^b_r]=i\varepsilon^{ab}{}_cQ^c_{m+r}+\tilde{\eta}^{ab}m\Psi_{m+r}
\label{eq4}\,,\\
&& [L_m,Q^b_r]=\left(\frac 12m-r\right)Q^b_{m+r}\label{eq5}\,,\\
&& \{Q^a_r,Q^b_s\}=2\tilde{\eta}^{ab}L_{r+s}
+i\varepsilon^{ab}{}_c(r-s)T^c_{r+s}
+\frac 13 \tilde{c}\left(r^2-\frac 14\right)\delta_{r+s,0}\tilde{\eta}^{ab}
\label{eq6}\,,\\
&& [L_m,\Psi_r]=\left(-\frac 12m-r\right)\Psi_{m+r}\label{eq7}\,,
\qquad
  [T^a_m,\Psi_r]=0\label{eq8}\,,\\
&& \{Q^a_r,\Psi_s\}=T^a_{r+s}\label{eq9}\,,\qquad
 \{\Psi_r,\Psi_s\}=\frac \ktSO 2 \delta_{r+s,0}\label{eq10}\,.
\end{eqnarray*}
The ${\cal N}=3$ superconformal algebra has a global subalgebra $OSp(3|2)$
whose bosonic part is associated with an isometry of
the $AdS_3\times S^3$.
Also super Lie algebra $OSp(3|2)$ 
is an $O(3)$ extended anti-de Sitter group in two dimension.
Modes of super stress tensors in space-time belong to vector
representation ${\bf 3}$ of this $O(3)\subset OSp(3|2)$. 
It is contrasted with the fact that the super stress tensors in 
the (small or large) ${\cal N}=4$ SCFT belong to spinor representation
of the $SU(2)\subset SU(1,1|2)$ or $SU(1,1|2)\times SU(1,1|2)$
respectively.

Let us comment on space-time physical operators.
We construct vertex operators associated to 
 $SL(2;\br)$\(\times\)$SU(2)^{(1)}$\(\times\)$SU(2)^{(2)}$.
We introduce following operators
\begin{eqnarray*}
 &&V_{j,m}=\gamma^{j+m}\exp \left[\frac{2j\phi}{\alpha_+}\right]\,,\\
 &&W_{j,m}\1=\frac{1}{\sqrt{(j+m)!(j-m)!}}
\gamma\1^{j+m}\exp \left[-i\frac{2j\phi\1}{{\alpha'}_+}\right]\,,\\
 &&W_{j,m}\2=\frac{1}{\sqrt{(j+m)!(j-m)!}}
\gamma\2^{j+m}\exp \left[-i\frac{2j\phi\2}{{\alpha'}_+}\right]\,,\\
&&\alpha_{+}=\sqrt{2 k_{s\ell}}\,,\,\,\,
{\alpha'}_{+}=\sqrt{2 k'}\,.
\end{eqnarray*}
The $V_{j,m}$ is the vertex operator in the $SL(2;{\br})$ algebra and
each $W_{j,m}^{(I)}$ $(I=1,2)$ belongs to a $K_0^{(I)3}=m$ state in the
spin \(j\)
representation of the affine $SU(2)^{(I)}$ algebra.
Then we present a vertex operator of the
space-time SCFT in the $(-1)$-picture
\[
 V=e^{-\varphi}(\psi^-\gamma+\psi^+\gamma^{-1}-2\psi^3)
V_{j,m}W_{j\p,m\p}\1W_{j\p\p,m\p\p}\2 \,.
\]
Its worldsheet conformal dimension $h$ is calculated as
\[
 h=\frac12+\frac12-\frac{1}{\kSL}j(j+1)
+\frac{1}{k\p}j\p(j\p+1)+\frac{1}{k\p}j\p\p(j\p\p+1).
\]
As a physical condition on the operator $V$, this \(h\) should be \(1\)
and we obtain a relation
\begin{eqnarray*}
  \frac{1}{\kSL}j(j+1)=\frac{1}{k\p}j\p(j\p+1)+\frac{1}{k\p}j\p\p(j\p\p+1).
\end{eqnarray*}
When one uses the level relation \(2\kSL=k\p\) (criticality
condition), 
it becomes an equation for the set $(j,j',j'')$
\[
  j(j+1)=\frac{1}{2}\left[\,j\p(j\p+1)+j\p\p(j\p\p+1)\,\right].
\]
We have to impose a ${\bz}_2$-invariant condition for the operators.
In this paper we restrict ourselves to the \(j\p=j\p\p\) case, for
simplicity. In this case, 
a relation \( j=j\p\;(=j\p\p)\) is automatically
satisfied, and the \({\bf Z}_2\)-invariant operators turn out to be
the spin \(2j\) representation of the SU\({}_D\)(2). Moreover,
 the highest weight state of this multiplet
 corresponds to a space-time primary state
with space-time conformal dimension $\tilde{h}=j$
and space-time $SU(2)$ spin \(\tilde{j}=2j\). This state with
 $2\tilde{h}=\tilde{j}$ represents a chiral primary
field of the space-time \(\Ncal=3\) theory.

\begin{figure}[htbp]
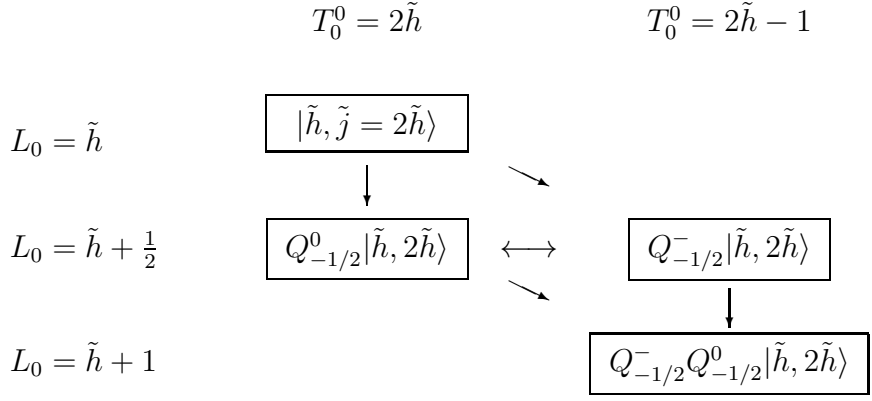

\begin{center}
  \begin{tabular}{lcccc}
   &\(T^0_0=2\tilde{h}\) &  & \(T^0_0=2\tilde{h}-1\) & \\
   & & & & \\
\(L_0=\tilde{h}\) &\framebox(76,21)
{ \(|\tilde{h},\tilde{j}=2\tilde{h}\rangle\) }
 & & &\\
           &\raise 5mm\hbox{\vector(0,-1){15}}
 &\raise 5mm\hbox{\vector(2,-1){15}} & & \\
\(L_0=\tilde{h}+\frac12\)\hspace*{10mm}&
 \fbox{ \(Q^0_{-1/2}|\tilde{h},2\tilde{h}\rangle\) } 
          &\(\longleftrightarrow\) &
\fbox{ \(Q^-_{-1/2}|\tilde{h},2\tilde{h}\rangle\) } \\
          & &\raise 5mm\hbox{\vector(2,-1){15}} 
&\raise 4mm\hbox{\vector(0,-1){15}}& \\
\(L_0=\tilde{h}+1\)& & &
\fbox{ \(Q^-_{-1/2}Q^0_{-1/2}|\tilde{h},2\tilde{h}\rangle\) } \\
 \end{tabular}
\end{center}
\caption{Short multiplet of the $\Ncal=3$ SCA. The state
 $|\tilde{h},\tilde{j}=2\tilde{h}\rangle $ is the chiral primary
 state. Other states in the multiplet are derived by multiplying $Q^-_{-1/2}$,
$Q^0_{-1/2}$} on the chiral primary.
\label{shortMultiplet}
\end{figure}

Now we recall that the short multiplet of the \(\Ncal=3\) SCFT
is constructed by multiplying \(Q^-_{-1/2},Q^0_{-1/2} \)
to the chiral primary state $\ket{\tilde{h}, \tilde{j}=2\tilde{h}}$
 as shown in figure
 \ref{shortMultiplet}.

Let us consider these space-time chiral primaries.
First we perform a standard twisting method on the
space-time ${\cal N}=3$ superconformal algebra.
All we have to do is to modify modes $(\tilde{L}_m,\tilde{Q}^0_r)$ of 
the energy momentum tensor and its ($SU_D(2)$ neutral ) super
partner as
\ba
&&\tilde{L}_m=L_m-\frac{1}{2}(m+1)T^0_m\,,\non
&&\tilde{Q}^0_r=Q^0_r-\left(r+\frac{1}{2}\right)\Psi_r\,,\non
&&\tilde{L}_0=L_0-\frac{1}{2}T^0_0\,.\label{twist}
\ea
This operation changes spins of charged currents under $SU(2)$
and the original ${\cal N}=3$ contents are rearranged into
four sets of fields in the table \ref{current}.
The mode of the 
current $Q^+_{r=-1/2}$ plays a role of a BRST operator $Q_{BRS}$
\ba
Q_{BRS}
=Q^+_{-1/2}\,,\nom
\ea 
and the modes of the energy momentum tensor are 
expressed in the BRST exact form
\ba
\tilde{L}_n=\frac{1}{4}\{Q_{BRS},Q^-_{n+\frac{1}{2}}\}\,.\nom
\ea
Similarly the currents $\tilde{Q}^0_r$, $Q^+_r$, $T^+_m$ are respectively
written as BRST exact forms of the $T^-_m$, $T^0_m$, $\Psi_r$.
Under this modification, only short multiplets are picked up as
physical states automatically because
states $Q^+_{-1/2}\ket{\cdots}=Q_{BRS}\ket{\cdots}$ are 
unphysical in that case.
\begin{table}[htbp]
 \begin{center}
  \begin{tabular}{|c||c|c||c|c|}\hline
 Spin  & $2$ & $3/2$ & $1$ & $1/2$ \\ \hline\hline
 Fields & \(\tilde{L}_m\) &\(\tilde{Q}^0_r\) & \(Q^+_r\) & $T^+_m$ \\ \hline
 Fields & \(Q^-_r\) &\(T^-_m\) & \(T^0_m\) & $\Psi_r$ \\ \hline
 \end{tabular}
 \end{center}
 \caption{Spins of the twisted ${\cal N }=3$ theory.
 When one modifies the $L_m$, $Q^0_r$ into the $\tilde{L}_m$,
 $\tilde{Q}^0_r$,
 the spins of the charged currents are changed by $\pm 1/2$.}
 \label{current}
\end{table}
Also two components of four sets $(\tilde{L}_m,\tilde{Q}^0_r)$,
$(Q^-_r,T^-_m)$, $(Q^+_r,T^+_m)$, $(T^0_m,\Psi_r)$ 
are connected each other by the super charge $Q_s$ with spin $1/2$
\ba
Q_s
=Q^0_{-1/2}\,.\nom
\ea
In particular, $\{Q_s,\tilde{Q}^0_{n+\frac{1}{2}}\}=2\tilde{L}_n$ is
satisfied.
This topological model has an ${\cal N}=1$ supersymmetry and
the states $\ket{\tilde{h},2\tilde{h}}$,
$Q^0_{-1/2}\ket{\tilde{h},2\tilde{h}}$
in the short multiplet are understood as a doublet of this susy algebra.
From the Eq.(\ref{twist}), we observe that 
an arbitrary state with $(L_0,T^0_0)=(\tilde{h},\tilde{j})$
is relabeled by a set of new eigenvalues of $(\tilde{L}_0,T^0_0)$
as $(\tilde{h}-\frac{1}{2}\tilde{j},\tilde{j})$.
In particular, original chiral primary state with $2\tilde{h}=\tilde{j}$
has eigenvalue zero for ${\tilde{L}_0}$ and is specified by the
eigenvalue  $\tilde{j}$ of ${T}^0_0$.
Also the state $Q^-_{-1/2}\ket{\tilde{h},2\tilde{h}}$
has dimension $1$ under the $\tilde{L}_0$ (with charge $2\tilde{h}-1$)
and integrated form of an associated scaling operator  
is interpreted as a marginal operator 
in space-time theory.
We may call the fields as moduli of some associated sigma model.

We propose these sets of fields as candidates of physical operators in 
some sigma model. 
Its target space might be
a symmetric product (Hilbert scheme) of some manifold.
We would like to discuss these subjects in future work.

Next we study this model in terms of worldsheet CFT and
make some speculations.
The zero-mode part of the $SU(2)$ current is constructed
by the two ${\cal N}=1$ affine $SU(2)$ currents in the
worldsheet
\ba
&&T^0_0=\oint \frac{dz}{2\pi i}K^3\,,\,\,\,
K^3=K^{(1)3}+K^{(2)3}\,,\non
&&K^{(I)3}=\beta^{(I)}\gamma^{(I)}+i\sqrt{k_{s\ell}}\del \phi^{(I)}
+\frac{1}{2}\chi^{(I)+}\chi^{(I)-}\,,\,\,\,(I=1,2)\,,\label{k3}
\ea
and the zero-mode of the modified stress tensor is 
expressed by the $SL(2;{\br})$ current together with ${K^3}$
\ba
&&\tilde{L}_0=\oint \frac{dz}{2\pi i}\left(-J^3-\frac{1}{2}K^3\right)\,,\non
&&J^3=\beta\gamma +i\sqrt{\frac{k_{s\ell}}{2}}\del \phi
+\frac{1}{2}\psi^+\psi^- \,.\label{j3}
\ea
The currents (\ref{k3}),(\ref{j3}) 
specify one complex structure of the (product of)
manifolds $AdS_3\times S^3_D\cong SL(2;{\br})\times SU_D(2)$
and the model is reduced to a coset type space
\ba
\frac{SL(2;{\br})_{k_{s\ell}}}{U(1)}\times 
\frac{SU_D(2)_{k_{so}}}{U(1)}\times U(1)^2\,.\nom
\ea
In terms of the worldsheet CFT, the currents $J^3$, $K^3$ are
parts of some $U(1)$ current $J$.
Also, it is well-known that
the $N=1$ superconformal algebra (\ref{n1}) is enhanced to
an $N=2$ CFT generated by $T$, $G=G^+ +G^-$ and $J$
in this coset case.
Now we concentrate on the $SU_D(2)_{k_{so}}/U(1)$ part.
It is a special case of the Grassmannian model
\ba
\frac{SU(n+m)_k}{SU(n)\otimes SU(m)\otimes U(1)}\,\,\,
\mbox{with}\,\,n=m=1\,,\,k=k_{so}\,.\nom
\ea
When we use the level-rank duality of the coset model,
this case is reduced to a level $1$ model
\ba
\frac{SU(2)_{k_{so}}}{U(1)}\rightarrow 
\frac{SU(k_{so}+1)_1}{U(k_{so})}\,.\nom
\ea
It is nothing but a $CP^{k_{so}}$-model 
with central charge $\dis c=3\left(1-\frac{2}{k_{so}}\right)$
and is described by a $A_{k_{so}+1}$ type minimal series
in the worldsheet $N=2$ theory or 2dim Landau-Ginzburg theory.
There could be a correspondence between operators of this $CP$-model and
those of $AdS_3$ model. 

Now we make several remarks here: the level ${k_{s\ell}}$ can be
interpreted as the square of the radius of the $AdS_3$.
In our case, it is related with the $k_{so}=4k_{s\ell}$ 
of the $SU(2)$ level and
$k_{so}$ must be integer. That is to say, the associated CFTs are
realized only on isolated integer points on the $k_{so}$ line.
When one fixes the $k_{so}$ as some number,
the associated CFT has at most finite number of primary fields
representing degenerate vacua of Landau-Ginzburg model.
However we can formally resolve the
degeneracy of the vacua by perturbation while preserving $N=2$ symmetry.
It is an interesting problem to clarify 
connections between the interpolating theory connecting different vacua 
and $AdS_3$ geometry.


\subsection{${\cal N}=1$ affine super Lie algebra}
Until now, we concentrate on the left-moving part of the worldsheet
theory. It generates a space-time ${\cal N}=3$
SCA in the left-handed side. In order to construct full theory, we
have to consider the remaining right-moving anti-holomorphic
part. The same recipes\footnote{We use the symbol `` $\bar{}$ '' (bar) 
to distinguish right-handed fields from the associated left-handed 
ones.} can be applied to this case in the bosonic parts.
Also one imposes conditions on right-moving spinors $\bar{u}(\bar{z})$
\ba
\Gamma_G\bar{u}=0\,,\,\,\,\Gamma_{\Omega}
\bar{u}=+\bar{u}\,,\nom
\ea
as physical conditions.
But we have to impose a negative chirality condition (GSO projection)
on the $\bar{u}$
\ba
\Gamma^{11}\bar{u}=-\bar{u}\,,\nom
\ea
because we consider the type \II A string. 
We summarize possible choices of
the GSO and \(\bz_2\) projections and associated
space-time supersymmetries in the table \ref{n3n1}.
\begin{table}[htbp]
\begin{center}
  \begin{tabular}{|c||c|c|}\hline
GSO \raise 10pt\hbox{\line(1,-1){10}} \(\bz_2\)
  &\(+\) &\(-\) \\ \hline\hline
\(+\) &\(\Ncal=3\) & \(\Ncal=1\) \\ \hline
\(-\) &\(\Ncal=1\) & \(\Ncal=3\) \\ \hline
 \end{tabular}
\end{center}
\caption{Choices of the GSO and \(\bz_2\) projections and
space-time superconformal symmetries. The signatures ($+$), ($-$)
for the GSO and ${\bz}_2$ projections mean
that the corresponding eigenvalues for these operators 
$\Gamma^{11}$, $\Gamma_{\Omega}$ are
$+1$, $-1$ respectively.}
\label{n3n1}
\end{table}

Here, we take a choice that
the GSO projection is `` \(-\) '' and \(\bz_2\)
projection is `` \(+\) '' case. Then we get the \(\Ncal=1\) SCA
in the right-handed side.
In fact these conditions select out only two possible spin operators
$\bar{S}_{\pm 1/2}$
\ba
 &&\Sb_{+1/2}=\frac{1}{2\sqrt2}
(+\bar{S}_{+-+++}+\bar{S}_{+-+--}+\bar{S}_{++-++}-\bar{S}_{++---})\,,\non
 &&\Sb_{-1/2}=\frac{1}{2\sqrt2}
(+\bar{S}_{--+-+}-\bar{S}_{--++-}+\bar{S}_{-+--+}+\bar{S}_{-+-+-})\,.\nom
\ea
They belong to a representation $({\bf 1},{\bf 2})$ of 
$SU_D(2)\times SL(2;{\br})$. In this case, a triplet of $SU_D(2)$ is
projected out and only ${\cal N}=1$ algebra is retained.
Then we can construct a set of modes of associated supercharge $\bar{Q}_r$
\ba
 \Qb_r=A \oint\frac{d\bar{z}}{2\pi i} e^{-\bar{\varphi}/2} 
\left[\left(r+\frac12\right)\gammab^{r-1/2}\Sb_{+1/2}
 -\left(r-\frac12\right)\gammab^{r+1/2}\Sb_{-1/2}\right]\,,\nom
\ea
in the space-time right-handed side. 
They satisfy an ${\cal N}=1$ superconformal algebra
\ba
 &&[\Lb_m,\Lb_n]=(m-n)\Lb_{m+n}+\frac{\ct}{12}(m^3-m)\delta_{m+n,0}\,,\non
 &&[\Lb_m,\Qb_r]=\left(\frac12m-r\right)\Qb_{m+r}\,,\non
 &&\{\Qb_r,\Qb_s\}=2\Lb_{r+s}+\frac{\ct}{3}\left(r^2-\frac14\right)
\delta_{r+s,0}\,,\nom
\ea
combined with Virasoro generators $\bar{L}_m$ in the right-moving side.
But the story is not completed yet. We have a degree of freedom of the
diagonal $S^3$, that is, the space-time $SU_D(2)$ affine symmetry.
We obtain modes of a set of the $SU_D(2)$ generators
$(\Tb^a_m,\Psib^a_r)$ \\
$(a=+,-,3\, ;\, r,s\in \frac{1}{2}+\bz\,;\,m\in\bz)$
\ba
&& \Tb^a_m=\oint \frac{d\bar{z}}{2\pi i}
\left[\gammab^m\Kb^a+2m\gammab^m\chib^a\left(
\psib^3-\frac12 \psib^-\gammab
-\frac12\psib^+\gammab^{-1}\right) \right]\qquad (0\,\mbox{-picture})\,,\non
 && \Psib^a_r=A\oint \frac{d\bar{z}}{2\pi i}
e^{-\bar{\varphi}/2}\bar{P}^a_r \qquad ((-1/2)\,\mbox{-picture})\,,
\non
&&\bar{P}^+_r=+\gammab^{r+1/2}\bar{S}_{-++++}
-\gammab^{r-1/2}\bar{S}_{+++-+}\,,\non
 && \bar{P}^0_r=
+\gammab^{r+1/2}\left(-\bar{S}_{--+-+}-\bar{S}_{--++-}
-\bar{S}_{-+--+}+\bar{S}_{-+-+-}\right)\non
&&\qquad\quad
+\gammab^{r-1/2}\left(+\bar{S}_{+-+++}
-\bar{S}_{+-+--}+\bar{S}_{++-++}+\bar{S}_{++---}\right)\,,\non
 && \bar{P}^-_r=-\gammab^{r+1/2}\bar{S}_{-----}
-\gammab^{r-1/2}\bar{S}_{+--+-}\,.
\nom
\ea
Thus space-time affine symmetry is a semi-direct product of the
${\cal N}=1$ super Virasoro and the ${\cal N}=1$ affine $SU(2)$ algebra.
They are generated by right-moving currents satisfying commutation
relations
\ba
&& [\Tb^a_m,\Tb^b_n]=i\varepsilon^{ab}{}_c \Tb^c_{m+n}
+\frac{\ktSO}{2}m\tilde{\eta}^{ab}\delta_{m+n,0}\,,\non
&& [\Tb^a_m,\Psib^b_r]=i\varepsilon^{ab}{}_c\Psib^c_{m+r}\,,\,\,\,
\{\Psib^a_r,\Psib^b_s\}=\frac{\ktSO}{2}\tilde{\eta}^{ab}\delta_{r+s,0}\,,\non
&& [\Qb_r,\Tb^a_m]=-m\Psib^a_{m+r}\,,\,\,\,
\{\Qb_r,\Psib^a_s\}=\Tb^a_{r+s}\,,\non
&& [\Lb_m,\Tb^a_n]=-n\Tb^a_{m+n}\,,\,\,\,
[\Lb_m,\Psib^a_r]=\left(-\frac12 m-r\right)\Psib^a_{m+r}\,.\nom
\ea
We collect these results in the table \ref{spacetimecft}.\\
\begin{table}[htbp]
 \begin{center}
 \begin{tabular}{|c||c|c|}\hline
 & \multicolumn{2}{c|}{\II A Theory}\\ \hline
 space-time & \makebox[50mm]{\hfil left-side\hfil} & right-side \\
 \hline
 chirality & $+$ & $-$\\
 \hline
 symmetry & (${\cal N}=3$ Virasoro) & \lower 14pt\hbox{\rule{0pt}{35pt}}
\parbox{50mm}{\baselineskip 13pt (${\cal N}=1$ Virasoro)\\
\hspace*{5mm} $\semidirectproduct$ (${\cal N}=1$ affine $SU(2)$)}\\
 \hline
 generators &\(L_m,\;Q^a_r,\;T^a_m,\;\Psi_r\)  & 
     \(\Lb_m,\;\Qb_r,\;\Tb^a_m,\;\Psib^a_r\)   \\ \hline
 \end{tabular}
 \end{center}
\caption{The structure of the space-time CFT. The chiral projections
  act on left-, right-sides with opposite signatures in the \II A
  string. It leads to generate a left-right asymmetric 
$({\cal N},\overline{\cal N})=(3,1)$ superconformal algebra.}
\label{spacetimecft}
\end{table}

\section{The brane configuration}
In this section, we study  a brane configuration realized by
M2-,M5-branes with a ${\bz}_2$-fixed plane in M-theory and 
show that it has a near horizon geometry 
\(AdS_3 \times(S^3 \times S^3 \times \br)/{\bz}_2 \times \br\).
If the last real line \(\br\) is compactified to a small circle \(S^1\),
the geometry becomes 
 \(AdS_3 \times(S^3 \times S^3 \times \br)/{\bz}_2\)
in the \II A theory 
considered in the previous section.

In the M5-M2 system, each brane extends to space-time directions shown in a 
table \ref{TableM5-O6-M2}.

\begin{table}[htbp]
\begin{center}
  \begin{tabular}{|c||c|c|c|c|c|c|c|c|c|c|c|}\hline
direction   &0 &1 &2 &3 &4 &5 &6 &7 &8 &9 &10 \\ \hline
 M5\({}\1\) &$\en$ &$\en$ &  &  &  &  &$\en$ &$\en$ &$\en$ &$\en$ & \\ \hline
 M5\({}\2\) &$\en$ &$\en$ &$\en$ &$\en$ &$\en$ &$\en$ &  &  &  &  & \\ \hline
fixed plane &$\en$ &$\en$ &/ &/ &/ &/ &/ &/ &/ &/ &$\en$ \\ \hline
 M2         &$\en$ &$\en$ &  &  &  &  &  &  &  &  &$\en$ \\ \hline
 \end{tabular}
\end{center}
\caption{M5-M2 system. Each brane extends 
to directions marked by circles. The symbol ``/'' (slash) means that
the fixed plane extends to the oblique directions. The M5\({}\1\) and
 the M5\({}\2\) are not independent sets of branes,
but one is the mirror image of the other.}
\label{TableM5-O6-M2}
\end{table}

The fixed plane extends to oblique directions. For convenience to express this 
fixed plane, we change coordinates from \(x\) to \(x\p\) 
\begin{eqnarray}
 &&x\p^p=x^p,\quad (p=0,1,10),\nonumber\\
 &&x\p^m=\frac{1}{\sqrt 2}(x^m+x^{m+4}),\quad (m=2,3,4,5),\nonumber\\
 &&x\p^{m\p}=\frac{1}{\sqrt 2}(x^{m\p-4}-x^{m\p}),\quad (m\p=6,7,8,9).
\nonumber
\end{eqnarray}
Then extended directions of 
the M5s and the M2s  are retained as in the table \ref{TableM5-O6-M2}
in old coordinates \(x\). But the fixed plane is stretched in
directions 0, 1, 2, 3, 4, 5, 10, in the new
coordinates \(x\p\).

\begin{figure}[htbp]
\epsfxsize=10cm
 \centerline{\epsfbox{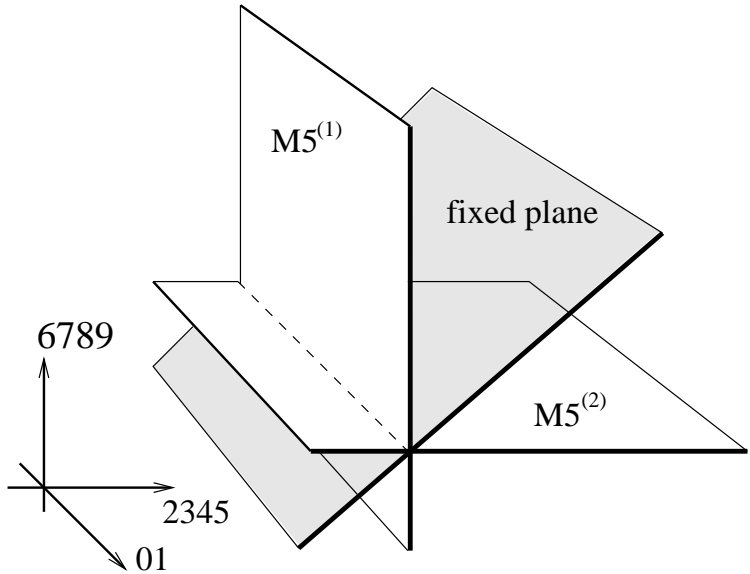}}
\epsfxsize=10cm
  \caption{M5-M2 system. The M2 is omitted. The M5\({}\2\) is the mirror
image of the M5\({}\1\) with respect to the fixed plane.}
\label{FigM5-O6-M2} 
\end{figure}

When we compactify the \(x^{10}\) direction to a small \(S^1\),
we get a configuration with NS5-NS1 branes in the \II A theory.

Now we will consider a classical solution associated with this 
brane system.
First we set the number of M5-branes
M5\({}\1\) ($\sharp$M5${}^{(1)}$) to be equal to that of M5\({}\2\)s 
($\sharp$M5${}^{(2)}$).
This set-up makes a ${\bz}_2$ action possible, which
exchanges M5\({}\1\)s and M5\({}\2\)s one another.

There is a metric of the classical solution in the  
M5\({}\1\)-M5\({}\2\)-M2 system\cite{sol} with 
$\sharp\mbox{M5}{}\1 =\sharp\mbox{M5}{}\2$
\begin{eqnarray}
 && ds^2=H_T^{1/3}H_{F1}^{2/3}H_{F2}^{2/3}\Big[
(H_TH_{F1}H_{F2})^{-1}\{-(dx^0)^2+(dx^1)^2\}\nonumber\\
 &&\qquad
+H_{F1}^{-1}\{(dx^2)^2+\cdots+(dx^5)^2\}
+H_{F2}^{-1}\{(dx^6)^2+\cdots+(dx^9)^2\}
+H_T^{-1}(dx^{10})^2
\Big],\nonumber\\
 && H_{F2}=1+\frac {R_5^2}{r_1^2}
,\qquad H_{F1}=1+\frac {R_5^2}{r_2^2}
,\qquad H_T= 1+\frac {R_2^4}{r_1^2r_2^2}\,,\nonumber\\
 && r_1^2=(x^2)^2+(x^3)^2+(x^4)^2+(x^5)^2
,\qquad r_2^2=(x^6)^2+(x^7)^2+(x^8)^2
+(x^9)^2\,.\nonumber
\end{eqnarray}
The \(R_5^2\), \(R_2^4\) 
are respectively constants proportional to the numbers of M5-branes \(Q_5\), 
M2-branes \(Q_2\).
This metric is invariant under following \({\bz}_2\) action 
\begin{eqnarray}
 x^m \to x^{m+4},\quad x^{m+4} \to x^m\qquad (m=2,3,4,5)\,.\nonumber
\end{eqnarray}
So we can divide this space by the \({\bz}_2\) and construct
a classical M5-M2 solution with a ${\bz}_2$-fixed plane.

Now let us take a near horizon limit and see what the geometry is.
The limit magnifies regions in the ranges 
\(r_1, r_2 \ll R_2\) and \(r_1, r_2 \ll R_5\) and then 
 the above metric becomes
\begin{eqnarray}
 && ds^2=\left(\frac{R^4}{r_1^2r_2^2}\right)^{-1}
\{-(dx^0)^2+(dx^1)^2\}\nonumber\\
 &&\qquad
+\frac {R^2}{r_1^2}dr_1^2+R^2(d\Omega_3\1)^2
+\frac {R^2}{r_2^2}dr_2^2+R^2(d\Omega_3\2)^2
+\left(\frac{R_5^4}{R_2^4}\right)^{2/3}(dx^{10})^2
\,.\nonumber
\end{eqnarray}
Here the \((d\Omega_3\1)^2\) and \((d\Omega_3\2)^2\) are the
metrics of two unit \(S^3\)s, and \(R\) is defined to 
be \((R_2^2R_5)^{1/3}\)
proportional to \((Q_2Q_5)^{1/6}\).
This metric is rather complicated and 
it is difficult to study the geometry directly.
To make the geometrical structure more clearly,
we perform a coordinate transformation \( (r_1,r_2) \to (u,v)\) by
\begin{eqnarray}
u=\ell \frac{r_1r_2}{R^2}\,,\qquad
v=\ell\log \left(\frac{r_1}{r_2}\right)\,, 
\qquad 2{\ell^2}={R^2}.\nom
\end{eqnarray}
Then we get a near horizon metric of this system,
\begin{eqnarray}
&& ds^2=\frac{\ell^2}{u^2}\{du^2-(dx^0)^2+(dx^1)^2\}
+R^2\{(d\Omega_3\1)^2+(d\Omega_3\2)^2\}+dv^2+
\left(\frac{R_5^4}{R_2^4}\right)^{2/3}(dx^{10})^2\,.
\nonumber
\end{eqnarray}
We find that the geometry 
is \(AdS_3\times S^3{}\1\times S^3{}\2 \times {\br}^2\) before
the ${\bz}_2$ operation.
 
When we treat M-theory as 11-dimensional classical supergravity,
we need to take the limit \(R\gg \ell_p\), where \(\ell_p\) is
the 11-dimensional Planck length. This is equivalent to a ``large N'' limit,
\[
 Q_2Q_5\gg 1.
\]

Also, in our solution, both radii of the two \(S^3\)s are \(R\).
The parameter $\ell$ is identified with a radius of the $AdS_3$.
They are connected by a relation 
\begin{eqnarray}
 2\ell^2=R^2\,.\label{ratio}
\end{eqnarray}
It relates a level
 of the worldsheet affine algebra \(\widehat{\rm SU(2)}_{k\p}\)
with that of the
 \(\widehat{\rm SL(2,R)}_{k_{s\ell}}\)
\begin{eqnarray}
 2k_{s\ell}=k\p.\nonumber
\end{eqnarray}

Now let us return to the \({\bz}_2\) action.
If we let \(\Omega\1^a\) and \(\Omega\2^a\), \((a=1,2,3)\) to be
coordinates of \(S^3{}\1\) and \(S^3{}\2\) respectively, 
the \({\bz}_2\) operation acts on them
\begin{eqnarray}
 \Omega\1^a \to \Omega\2^a,\quad \Omega\2^a \to \Omega\1^a,\quad 
 v \to -v.\nom
\end{eqnarray}
Here we find that the \({\bz}_2\) action considered in the previous
section is a natural operation. 
After the division, we obtain a resulting space
 \(AdS_3\times ( S^3 \times S^3 \times {\br})/{\bz}_2 \times {\br}\). 
Because the \(x^{10}\) direction is flat in this geometry,
we can compactify it
to a small circle \(S^1\). 
By decreasing the radius of the $S^1$, the geometry in the M-theory
is reduced to the
\(AdS_3\times ( S^3 \times S^3 \times {\br})/{\bz}_2\) in  
the type \II A theory.
It is just the same theory that we consider in the previous section.

Finally we comment on the fixed plane. On this fixed surface,
next relations are satisfied
\begin{eqnarray}
 && \Omega\1^a=\Omega\2^a=\Omega_3^a,
\quad v=0,\nonumber\\
 && d\Omega\1^a=d\Omega\2^a=d\Omega_3^a
,\quad dv=0\,.\nom
\end{eqnarray}
The geometry of this surface is characterized as 
\(AdS_3\times S^3 \times {\br}\) and its induced metric is given as
\begin{eqnarray}
  ds^2=\frac{\ell^2}{u^2}\{du^2-(dx^0)^2+(dx^1)^2\}
+2 R^2(d\Omega_3)^2
+\left(\frac{R_5^4}{R_2^4}\right)^{2/3}(dx^{10})^2\,.\nom
\end{eqnarray}
A radius of \(S^3\) part is understood to be 
\(\tilde{R}=\sqrt2 R\). 
By comparing the radius $\ell$ of $AdS_3$ with this $\tilde{R}$,
we obtain a relation between them.
They are connected with an equation
\begin{eqnarray}
 4\ell^2=\tilde{R}^2\,.\nom
\end{eqnarray}
It gives us a ratio of two levels 
of the worldsheet \(\widehat{\rm SL(2,{\bf R})}_{k_{s\ell}}\) and the 
diagonal \(\widehat{\rm SU_D(2)}_{k_{so}}\) as
\begin{eqnarray}
\frac{k_{so}}{k_{s\ell}}=\frac{\tilde{R}^2}{\ell^2}=4\,.\nom
\end{eqnarray}
It coincides with the result derived in the previous section.

In this brane configuration, the M2-brane ends on the M5s.
The intersecting part is a (1+1) dimensional object and represents a 
string. When one compactifies one longitudinal dimension,
it is known as a little string\cite{little} trapped in the
NS5-brane in the context of the small iia theory.
The associated low energy theory is believed to be an
exotic 6dim (2,0) type \II A susy theory.
In our set-up, our susy model lives on a fixed plane
with these M2-M5 backgrounds.
We might have a new low energy susy theory
with some additional multiplets.
We hope to study these and clarify a description of them in the
M(atrix) or Quiver matrix theory in future works.

\section{Conclusions and Discussions}
\cleqn

In this paper, we investigated the superstring on the background 
$AdS_3\times( S^3\times S^3\times S^1)/{\bz}_2$ in the
framework presented in \cite{GKS} and developed a method to construct 
a conformal theory with less supersymmetry in space-time.
We study a WZW model on $AdS_3$ target space with a ${\bz}_2$-action 
and find that an $SL(2;{\br})\times SU_D(2)$ isometry is enhanced to the
${\cal N}=3$ superconformal algebra in space-time.
Global symmetry part is a super Lie algebra $OSp(3|2)$ 
which is an $O(3)$ extended anti-de Sitter group in two dimension.

The criticality condition in this superstring is equivalent to a
relation $k_{so}=4k_{s\ell}$
of two levels $k_{so}$, $k_{s\ell}$ associated with affine Lie algebras
$SU_D(2)$ and $SL(2;{\br})$ respectively.
It also gives us information that a ratio among the radii of the
diagonal $S^3_D$ and $AdS_3$ is two.

In order to construct supercharges (spin operators), we used standard
bosonization techniques for worldsheet fermions. We imposed three
constraints on these spin operators as physical conditions.
The first is the usual BRS condition. 
The second is reduced to the ${\bz}_2$-invariance condition.
The ${\bz}_2$-operation acts on two $S^3$'s and exchanges their
associated fields one another. 
Under the ${\bz}_2$-projection, a diagonal part of products of two
$S^3$'s is survived and it serves as an R-symmetry of the space-time
theory. 
It also changes signatures of fields
on the circle $S^1$.
The last condition comes from a GSO projection. 

When one imposes a positive chirality condition, only spinor
representation is survived. Oppositely co-spinor operators are 
obtained under a negative chirality constraint.
We have four possible choices for these ${\bz}_2$-
and GSO projections, that is, $(+,+)$, $(-,-)$, $(+,-)$, $(-,+)$. 
The choices $(+,+)$, $(-,-)$ lead to ${\cal N}=3$
superconformal algebras in space-time.
These spin fields ${S^a_r}$ belong to representation  
$({\bf 2},{\bf 3})$ in the $SL(2;{\br})\times SU_D(2)$.
But in the remaining two cases, possible spin operators
belong to $({\bf 2},{\bf 1})$ representation of
$SL(2;{\br})\times SU_D(2)$ and only ${\cal N}=1$ algebras are
retained. They satisfy ${\cal N}=1$ superconformal
algebra (SCA) combined with Virasoro generators.

In the context of the \II A string theory, the chiralities 
of supercharges 
are opposite in space-time and the GSO projections act on
left-, right-moving parts with different signatures.
The resulting type \II A theory has an 
$({\cal N},\overline{\cal N})=(3,1)$ superconformal symmetry in
space-time
with an extra affine $SU(2)$ algebra in the $\overline{\cal N}=1$ part.
The $\overline{\cal N}=1$ part
has a degree of freedom of the diagonal 
$S^3_D$, that is, affine $SU_D(2)$ symmetry in space-time.
We construct modes of a set of these $\overline{\cal N}=1$ affine $SU_D(2)$ 
generators explicitly.
The precise
physical meaning of this remains to be understood. 

Also we present physical vertex
operators, in particular, chiral primary fields in the 
${\cal N}=3$ SCFT. 
We study short multiplets of (chiral part of) the ${\cal N}=3$ theory.
It has ${\cal N}=1$ doublet states and
we identify (integrated form of)
$Q^-_{-1/2}\ket{\tilde{h},2\tilde{h}}$s 
with marginal operators. 
A candidate for the space-time $({\cal N},\overline{\cal N})=(3,1)$
full theory is a orbifold-type sigma model like some symmetric product
(more precisely Hilbert schemes) of a compact space.
In order to establish the $AdS_3/CFT_2$ correspondence, we have to
consider the total Fock space of the space-time CFT and clarify
the properties of the space-time vacua.

Next, from the point of view of worldsheet theory,
we speculate about the roles of the currents $J^3$, $K^3$ and
about relations between $N=2$ $CP^{k_{so}}$-model and
$AdS_3$-model.
There are several integrable deformations in the context of the CFT.
For an example, we can deform the theory by the most relevant
operator associated with a K\"ahler form of the $CP$ model.
As another example, one can obtain Toda-type theories
by using Chebyshev polynomials and then their solitons 
connect different vacua.

In the viewpoint of M-theory, we present a brane configuration whose
near horizon geometry describes the $AdS_3$ background.
The geometry is obtained at the throat limit of 
two sets of M5-branes intersecting in one direction,
together with M2-branes with a fixed-plane
extending to oblique directions.
The two kinds of M5-branes (M5${}^{(1)}$, M5${}^{(2)}$)
are mirror images one another. 
Then the ${\bz}_2$-group acts on a classical solution of the
M2, M5${}^{(1)}$, M5${}^{(2)}$ system. By taking a large charge limit,
we can obtain a near horizon geometry
$AdS_3\times (S^{3(1)}\times S^{3(2)}\times {\br})/{\bz}_2 \times {\br}$
with a natural ${\bz}_2$ action
($S^{3(1)}\leftrightarrow S^{3(2)}$, ${\br}\leftrightarrow -{\br}$ ).
By shrinking the radius of a longitudinal direction $\br$, we 
construct the orbifold-type space with the ${\bz}_2$-action in the \II A 
theory. 

The M2 brane ends on the M5
and the intersecting part is a $(1+1)$ dimensional object.
When one compactifies one longitudinal direction (in stringy limit), 
it is known as a
little string trapped in the NS5-brane.
In our set-up, conformal theory lives on the fixed plane
in the background with solitonic NS5-branes and little strings.
We might have a new exotic low energy susy gauge theory with some
extra multiplets.
The near horizon region in this model is the same as an infrared
limit for this gauge theory.
It is a challenging problem to analyse these in the
M(atrix) model. 

In this paper, for simplicity, we consider only a ${\bz}_2$-action.
However we would expect that the method developed here is applicable
to other general backgrounds with actions of discrete groups.
We would like to discuss these subjects elsewhere.

\section*{Acknowledgement}
K.~S. and S.~Y. are grateful to T.~Uematsu for valuable
discussions and useful comments. 
Y.~I. gratefully acknowledges helpful conversations with
S.~Matsuda. The authors also 
thank the Yukawa Institute for Theoretical
Physics for hospitality and the participants of the
associated seminars for discussions.
K.~S. is supported in part by the Grant-in-Aid for
Scientific Research from the Ministry of Education, Science, Sports
and Culture 10740117.
S.~Y. is a Research Fellow of
the Japan Society for the Promotion of Science 
and is supported in part by the Grant-in-Aid
for Scientific Research from the Ministry of Education, Science, Sports
and Culture.

\newpage
\appendix
\setcounter{section}{1}
\section*{Appendix A. SO(10) Gamma matrices}
\cleqn
Here we summarize the SO(10) Gamma matrices.

We choose a convention for Gamma matrices \(\Gamma^M,\;(M=1,\cdots,10)\)
\begin{eqnarray*}
 &&\Gamma^1=\sigma^1\otimes \sigmao \otimes \sigmao \otimes \sigmao 
\otimes \sigmao   \,,\nonumber\\
 &&\Gamma^2=\sigma^2\otimes \sigmao \otimes \sigmao \otimes \sigmao 
\otimes \sigmao \,,\nonumber\\
 &&\Gamma^3=\sigma^3\otimes\sigma^1\otimes \sigmao \otimes \sigmao 
\otimes \sigmao \,,\nonumber\\
 &&\Gamma^4=\sigma^3\otimes\sigma^2\otimes \sigmao \otimes \sigmao 
\otimes \sigmao \,,\nonumber\\
 &&\Gamma^5=\sigma^3\otimes\sigma^3\otimes\sigma^1\otimes 
\sigmao \otimes \sigmao \,,\nonumber\\
 &&\Gamma^6=\sigma^3\otimes\sigma^3\otimes\sigma^2\otimes 
\sigmao \otimes \sigmao \,,\nonumber\\
 &&\Gamma^7=\sigma^3\otimes\sigma^3\otimes\sigma^3\otimes
\sigma^1\otimes \sigmao \,,\nonumber\\
 &&\Gamma^8=\sigma^3\otimes\sigma^3\otimes\sigma^3\otimes
\sigma^2\otimes \sigmao \,,\nonumber\\
 &&\Gamma^9=\sigma^3\otimes\sigma^3
\otimes\sigma^3\otimes\sigma^3\otimes\sigma^1\,,\nonumber\\
 &&\Gamma^{10}\!\!=\sigma^3\otimes\sigma^3
\otimes\sigma^3\otimes\sigma^3\otimes\sigma^2\,,
\end{eqnarray*}
where the $\sigma^a$, $(a=1,2,3)$ are the Pauli matrices, and \(\sigmao\) is
the \(2\times 2\) unit matrix described as
\begin{eqnarray*}
\sigma^1=\left(
 \begin{array}{cc}
 0 & 1\\
 1 & 0\\
 \end{array}\right),\;
\sigma^2=\left(
 \begin{array}{cc}
 0 & -i\\
 i & 0\\
 \end{array}\right),\;
\sigma^3=\left(
 \begin{array}{cc}
 1 & 0\\
 0 & -1\\
 \end{array}\right),\;
\sigmao=\left(
 \begin{array}{cc}
 1 & 0\\
 0 & 1\\
 \end{array}\right).\;
\end{eqnarray*}
These \(\Gamma^M\)'s satisfy 
anti-commutation relations
 $\{\Gamma^M,\Gamma^N\}=2\delta^{MN}$.
The chirality operator \(\Gamma^{11}\) is expressed as
 \begin{eqnarray*}
  && \Gamma^{11}=-i\Gamma^1\Gamma^2\cdots\Gamma^{10}
   = \sigma^3\otimes \sigma^3\otimes \sigma^3\otimes
 \sigma^3\otimes \sigma^3 .
 \end{eqnarray*}
Also we define a unitary matrix $C$ representing a
charge conjugation operation as 
\begin{eqnarray*}
   C=\sigma^2\otimes\sigma^1\otimes\sigma^2\otimes
\sigma^1\otimes\sigma^2\,.
\end{eqnarray*}
This \(C\) satisfies following several relations 
\begin{eqnarray*}
 C\Gamma^MC^{-1}=-(\Gamma^M)^T,\quad C=C^{\dagger}=C^{-1}=-C^T.
\end{eqnarray*}
Then the \(C\Gamma^M\) is symmetric as a matrix.

In this paper, these representations of Gamma matrices appear as 
cocycle factors in the
the following OPEs including spin fields
\begin{eqnarray*}
&& \psb^M(z)S_{\alpha}(w) = \frac{1}{\sqrt 2 (z-w)^{1/2}}
S_{\beta}(\Gamma^M){}^{\beta}{}_{\alpha}\,,\non
 && e^{-\varphi/2}S_{\alpha}(z) e^{-\varphi/2}S_{\beta}(w)\sim 
\frac{1}{z-w}e^{-\varphi}(C\Gamma^M)_{\alpha\beta}\psb^M\,.\nom
\end{eqnarray*}

\end{document}